\documentstyle[12pt]{article}                                           
  \def\ocomma{\buildrel\otimes\over ,}                            
\begin{document}                                             
  \rightline{HUP-96/001 }                                       
  \rightline{DTP-96/27 }
  \rightline{hep-th/9612150}
    
  \begin{center}
  {\Large Classical integrability of the $O(N)$ nonlinear \break
Sigma model on a half-line } 
  \vskip 1.1cm
  \centerline{  
  E. Corrigan$^{a}$ and Z-M Sheng$^{b}$\footnote{email : 
  zmsheng@whale.hzuniv.edu.cn}}
  \vskip 0.8cm
   { \it a. Department of Mathematical Science, University of Durham, \\[2mm] 
   Durham, DH1 3LE, U.K. 
   \\[2mm]
   b. Department of Physics, Hangzhou University,\\[2mm]
  ~~~~Hangzhou 310028, P.R. China }
  \vskip 1.0cm
  {\sc Abstract }
  \vskip 0.7cm
    
  \begin{minipage}{12cm}
The classical integrability the 
$O(N)$ nonlinear sigma model on a half-line is examined, and the
existence of an infinity of conserved charges in involution is
established for the free boundary condition. 
For the case $N=3$ other possible boundary conditions are 
considered briefly.    
  \end{minipage}
  \end{center}
  \vskip 1.0cm
  July 1996
  \vfill
  \eject
    
  \noindent{\bf Introduction}
    
  Recently, much attention has been paid to the study of integrable 
  field theory with a boundary 
\cite{Skl}\cite{GZ}\cite{Gh1}\cite{CDRS}\cite{BCDR}\cite{SS}\cite{PZ}\cite{SG}. 
  The study of such  field theories is not only  intrinsically interesting 
   but also provides a better understanding of boundary related 
  phenomena in statistical physics and condensed matter  \cite{LeC}\cite{Wen}. 
  Probably the most famous physical example of a boundary integrable model 
  is the Kondo problem, where the 1+1 dimensional field theory is an 
  effective 
  field theory of s-wave scattering of electrons off a magnetic spin 
  impurity. Such an impurity problem, in which one concentrates on s-wave 
  scattering from some isolated object at the origin, generically provides 
  interesting 1+1 dimensional boundary field theory. There are also 
  1+1 dimensional boundary quantum integrable systems of 
  experimental relevance, such as a Luttinger liquid (Thirring) model for 
  the edge states of electrons in the fractional quantum Hall 
  effect  \cite{Wen}.
    
  An integrable field theory possesses an infinite set of independent, 
commuting integrals of motion. In the \lq bulk theory' these integrals 
of motion follow from an 
infinite number
of divergenceless currents. However, when the theory is restricted to a 
half-line  (or to an interval) the 
existence of conserved charges on the whole-line does not 
guarantee integrability unless special boundary conditions are specified. 
The boundary conditions under which 
  a theory preserves its integrability can be sought in several
  ways: for example,  via a perturbed conformal boundary 
  condition  \cite{Cardy}\cite{CL} or via a Lax pair 
  approach \cite{BCDR}.
    
  However, it is more difficult to discuss the integrability for 
  non-ultralocal integrable models, such as a nonlinear sigma model on
  the half-line, because
  their fundamental Poisson bracket relations are field dependent 
\cite{Pol}\cite{LP}\cite{Mail}\cite{BFLS}\cite{HHLW}\cite{GT}. 
  Ghoshal\cite{Gh2} has obtained the boundary S-matrix of the 
  O(N)-symmetric nonlinear sigma model, when he conjectured that this model 
  is still integrable in the semi-infinite space with the \lq free' or 
  \lq fixed' boundary conditions. However, Mourad and Sasaki\cite{MS} 
 have concluded recently
   that the infinite set of non-local charges characterizing  
  integrability on the whole line is not preserved even for the free (Neumann)
  boundary condition.  In this paper, the question of integrability for
  sigma models with a boundary will be re-examined from a different view-point.
  First, standard tools for the 
  model on the full line will be reviewed. In particular, 
  the Lax pair representation, the $r,~s$ matrices, and their 
  use to generate conserved charges, will be discussed in the next section. 
  Following the ideas of \cite{BCDR}, a 
  Lax pair on the half-line will be constructed in the third section,
  with the property that its zero curvature condition leads to both the 
  equation of motion and to the boundary condition. In the fourth section, 
  the Lax pair will be used to construct 
  the conserved charges, suggesting integrability of the model 
  on the half-line, and to determine compatible 
integrable boundary conditions.          
    
  \noindent{\bf The $O(N)$ nonlinear sigma model on the full line}
    
  To establish notation,  the standard Lax pair for the $SO(N)$ nonlinear 
  sigma 
  model \cite{Pol} will be written in the form
  \begin{equation}
  \begin{array}{ll}
  a_0 &= {\lambda \over 1 - \lambda^2}(\lambda A_0 + A_1) \\[3.mm]
  a_1 &= {\lambda \over 1 - \lambda^2}(\lambda A_1 + A_0)
  \end{array} 
  \end{equation}
  where 
  \begin{equation}
  \begin{array}{ll}
  A_{\mu} = 2 (n\otimes\partial_{\mu}n 
   -  \partial_{\mu} n\otimes n), \\[2mm]
  \end{array} 
  \end{equation}
  and $n$ is an $N$-component, unit length vector.
    
  It is easy to check that the zero curvature condition for the
  vector potentials defined by eq(1),
  \begin{equation}
  \partial_0 a_1 -  \partial_1 a_0 + [a_1,  a_0] = 0,
  \end{equation}
  leads to the equation of motion of the  model
  in the form
  \begin{equation}
   n\otimes\partial^2 n-\partial^2 n\otimes n=0,
  \end{equation}
  or, equivalently, 
  \begin{equation}
  \partial^2 n_a + n_a (\partial_{\mu}n_b \partial_{\mu}n_b) = 0.
  \end{equation}
  The action of the $O(N)$ nonlinear sigma model may be taken to be
  \begin{equation}
   S= \frac{1}{2}\int d^{2}x\left( (\partial_{\mu} n_a)^2 + \zeta 
( n_a^2 -1) \right),
  \end{equation}
where $\zeta$ is a Lagrange multiplier.
  The canonical momentum associated to $n_a$ is $\pi_a = \partial_0 n_a$.  
  However,  the Hamiltonian formalism is of 
  constrained type. The Hamiltonian is
  \begin{equation}
   H = \frac{1}{2} \int dx [ \pi_a^2 +(\partial_1 n_a)^2 ]
  \end{equation}
  and the  canonical Poisson brackets are
  \begin{equation}
  \begin{array}{l}
  \{ n_a(x),~ n_b(y) \} =0 \\[2mm]
  \{ n_a(x),~ \pi_b(y) \} = (\delta_{ab} -n_{a}n_{b})\delta (x-y) \\[2mm]
  \{ \pi_a(x),~ \pi_b(y) \} = (\pi_{a}n_b - \pi_b n_a)\delta (x-y)
  \end{array}
  \end{equation}
It is also useful to introduce a representation for the $SO(N)$ generators
\begin{equation}
T^{ab}_{\alpha\beta}=\delta^a_\alpha\delta^b_\beta -\delta^a_\beta
\delta^b_\alpha ,
\end{equation}
with quadratic Casimir 
\begin {equation}
C={1\over 2}T^{ab}\otimes T^{ab},
\end{equation}
and to set, using the components (2),
\begin{equation}
A_\mu =2n_a\partial_\mu n_b\, T^{ab}.
\end{equation}
It is also convenient to define an additional quantity
\begin{equation}
A=n_aT^{ab}\otimes T^{bc}n_c.
\end{equation}
The two operators $C$ and $A$ play a significant role in setting up 
the Poisson structure of the model. The use of the same notation for
the matrices, or components should cause no confusion.
 
Using the Poisson brackets (8), and following closely Bordemann 
et al. \cite{BFLS}, one may obtain the following algebraic structure:
  \begin{equation}
  \begin{array}{l}
  \{ A_0(x) \ocomma A_0(y) \} = 2 [C, ~I\otimes A_0]\delta(x-y)\\[2mm]
  \{ A_1(x) \ocomma A_1(y) \} = 0 \\[2mm]
 \{ A_0(x) \ocomma A_1(y) \} = 2 [C, ~I\otimes A_1]
 \delta(x-y) - 4 A(y)\delta^\prime (x-y). \\[2mm]
 \end{array}
 \end{equation}
For the last of these, use has been made of the identity
$$ ( f(x) - f(y)) \delta^{\prime}(x-y)= - f^{\prime}(x)\delta (x-y).$$

In turn, eqs(13) may be used to calculate the Poisson brackets of the
Lax pair operators themselves, casting them into the Maillet form:
 \begin{eqnarray}
 &&\{ a_1(x, \lambda) \ocomma a_1(y, \mu) \}
 =\Bigl([ r(x,\lambda,\mu)~, ~a_1(x, \lambda) \otimes I 
  + I \otimes a_1(x, \mu)] 
 \nonumber\\[2mm]
  &&\phantom{\{ a_1(x, \lambda) \ocomma a_1(y, \mu}\ \ -[ s(x,\lambda,\mu)~,~ 
    a_1(x, \lambda)\otimes I -I\otimes a_1(x, \mu) ]\Bigr)
\delta(x-y) \nonumber \\[2mm]
  &&\phantom{\{ a_1(x, \lambda) \ocomma } \ + 
\Bigl( r(x,\lambda,\mu) + s(x, \lambda, \mu) - 
r(y, \lambda, \mu) 
  + s(y, \lambda, \mu)\Bigr)\delta^{\prime}(x-y) 
 \end{eqnarray}
where
 \begin{eqnarray}
&&  r(x,\lambda,\mu)= -\frac{2 \lambda\mu}
  {(\lambda -\mu)(1 - \lambda \mu)}
  C + \frac{2\lambda \mu (\lambda - \mu)(1 + \lambda\mu)}
   {(1 - \lambda^2)(1 - \mu^2)(1 - \lambda \mu)}A (x) \\[2mm]
&&  s(x,\lambda,\mu) = 
 \frac{2\lambda\mu (\lambda + \mu)}
  {(1- \lambda^2)(1- \mu^2)}A(x).
 \end{eqnarray}
In checking eqs(14)  the following additional relationship is
indispensable
\begin{equation}
[A,~ I\otimes A_{\mu} - A_{\mu} \otimes I] = [C,~A_\mu \otimes I ]
 \end{equation}
  
In equations (14), $r(x,\lambda,\mu)$ and $s(x,\lambda,\mu)$ are two 
  matrix \lq structure 
  constants' which  depend on $x$ because they depend on the field of the 
  theory.
They have following property
 \begin{equation}
  r(x,\lambda,\mu) = - r(x, -\lambda -\mu)
 \end{equation}
  and satisfy the following modified Yang-Baxter equation
 \begin{equation}
 \begin{array}{l}
  \Bigl[ (r + s)_{23}(x, \mu, \eta)~,~ (r + s)_{12}(x, \lambda, \eta)\Bigr]
    + \Bigl[ (r + s)_{23}(x, \mu, \eta)~,~ (r + s)_{13}(x, \lambda, 
\eta)\Bigr]\\[2mm]
 + \Bigl[(r + s)_{13}(x,\lambda, \eta) ~,~ (r - s)_{12}(x, \lambda, \mu)\Bigr] 
  + H^{(r+s)}_{1,23}(x, \lambda, \mu, \eta) - H^{(r+s)}_{2,13}
(x, \mu, \lambda, \eta ) = 0 ,
 \end{array}
 \end{equation}
  where  $ H^{r + s}_{1,23}(x, \lambda, \mu, \eta)$
is defined by
 \begin{equation}
 \{a_1(x, \lambda) \otimes I~,~ (r +s)_{23}(y,\mu,\eta)\} 
  = H^{r + s}_{1,23}(x, \lambda, \mu, \eta) \delta(x-y).
 \end{equation}
  
  One of the attractive features of the Lax pair representation 
is the role it plays
  in the generation of conserved quantities. 
  The path-ordered exponential
 \begin{equation}
  T(x,y; \lambda) = {\rm P} \exp \int_{x}^{y} a_1(z, \lambda) dz
 \end{equation}
  enjoys the following properties,
 \begin{equation}
 \begin{array}{l}
  T(x,x;\lambda) = 1,~~~ T(y,x;\lambda) = T^{-1}(x,y; \lambda),\\[2mm]
  \, \, \, \, \ \  T(x,y;\lambda) T(y,z; \lambda) = T(x,z; \lambda),
 \end{array}
 \end{equation}
  and satisfies
 \begin{equation}
 \begin{array}{l}
 \{ T(x,y; \lambda) \ocomma T(x^{\prime},y^{\prime}; \mu) \} \\[2mm]
 \ \ \  =\int^{y}_{x} dz \int^{y^{\prime}}_{x^{\prime}} dz^{\prime} T(x,z; 
 \lambda)\otimes T(x^{\prime},z^{\prime}; 
 \mu) \{a_1(z,\lambda) ~\ocomma~ a_1(z^{\prime}, \mu) \} T(z,y, \lambda) 
 \otimes T(z^{\prime},y^{\prime}; \mu).
 \end{array}
 \end{equation} 
  Defining $ U(\lambda)
  = T( -\infty, +\infty; \lambda)$, 
  one obtains from eq(23) the relation
 \begin{equation}
 \{ U(\lambda) \ocomma U(\mu) \} = r(- \infty, \lambda, \mu)U(\lambda) 
 \otimes U(\mu) - U(\lambda) \otimes U(\mu) r( +\infty,\lambda, \mu). 
 \end{equation}
  Alternatively, assuming  periodic boundary conditions on the finite 
interval $(-L,L)$, ie $n_a( -L) = n_a(+L)$, one has
 \begin{equation}
 \begin{array}{l}
 \{ U_L(\lambda) \ocomma U_L(\mu) \} = [ r(- L, \lambda, \mu)~,~ 
  U_L(\lambda) \otimes U_L(\mu)] \\ [2mm]
 \ \ \  + ( U_L(\lambda) \otimes I) s (L,\lambda,\mu) (I \otimes U_L(\mu)) 
  - ( I \otimes U_L(\mu)) s (L,\lambda,\mu) (U_L(\lambda) \otimes I), 
 \end{array}
 \end{equation}
  where $U_L(\lambda) = T(-L, L, \lambda)$. From eq(24) or eq(25), 
it is straightforward to obtain 
 \begin{equation}
 \{ trU(\lambda)~, trU(\mu) \}=0
 \end{equation}
  indicating that the infinity of conserved charges obtained from 
  $U(\lambda)$
  as coefficients in an expansion in powers of $\lambda$ are in involution.
  
\bigskip
  
 \noindent{\bf Lax pair representation for O(N) sigma model on a half-line}
  
  A Lax pair on a half-line will be constructed so that its zero 
  curvature condition  leads both to the field  equation and to
  the boundary condition at $x=a$. The procedure developed in \cite{BCDR}
  will be followed closely.
  To construct a modified Lax pair including the boundary condition, it is 
  first of all convenient to consider an additional special point $x = b 
 \ ( > a)$  and two overlapping regions $ R_{-} :~ x \leq 
  (a+b+\epsilon)/2$; and $R_{+} :~ x \ge (a+b-\epsilon)/2$. The 
  second region 
  will be regarded as a reflection of the first, in the sense that if 
  $x\in R_{+}$, then
 \begin{equation}
  n_a(x) = n_a(a+b-x).
 \end{equation}
  The regions overlap in a small interval surrounding the midpoint of 
  $[ a,b]$. In the two regions, define:
 \begin{equation}
 \begin{array}{ll}
  R_{-}: ~~~ &\hat{a}_0 = a_0 + \frac{2\lambda}{(\lambda^2-1)}\theta(x-a)
 \left[ n\otimes \left(\partial_1 n + 
{\cal F}\right) -\left(\partial_1 n + 
{\cal F}\right)\otimes n\right] \\[2mm]
  &\hat{a}_1 =\theta(a - x) a_1 \\[2mm] 
  R_{+}: ~~~ &\hat{a}_0 = a_0 + \frac{2\lambda}{(\lambda^2-1)}\theta(b-x)
 \left[ n\otimes \left(\partial_1 n - 
{\cal F}\right) -\left(\partial_1 n - 
{\cal F}\right)\otimes n\right] \\[2mm]
  &\hat{a}_1 = \theta(x - b) a_1 
 \end{array}
 \end{equation} 
  It is then straightforward to check this  Lax pair leads to the  
equation 
  of motion on the 
  half-plane, and to the boundary conditions at the boundary point $x = a$:
 \begin{equation}
 \partial_1 n_a = - {\cal F}_a.
 \end{equation}
For the moment, the question of the form of the boundary potential term
which ought to be added to the Lagrangian will be postponed. 
  
  On the other hand, for $x \in R_{-} $ and $ x > a$, $\hat{a}_1$ vanishes 
  and therefore the zero curvature condition merely implies 
  that $\hat{a}_0$ 
  is independent of x, which means in turn that the fields $n_a$ are 
  independent of $x$ in this 
  region. Similar remarks apply to the region $x \in R_{+}$ and $x < b$. 
  Hence, taking into account the reflection principle, fields are 
  independent of $x$ throughout the interval $[ a ,b ]$, and equal to the
  field value 
   at $a$ or $b$. For general boundary conditions, the gauge 
  potential   $\hat{a}_0$ is different in the two regions $R_{\pm}$. 
  However, to maintain  the zero curvature condition over the 
  whole line the values of $\hat{a}_0$
  must be related by a gauge transformation within the overlap region $[a,b]$. 
  In principle, this gauge transformation could be quite general, 
even depending upon the field $n$.
 Here, it will be assumed 
 a gauge transformation $\kappa(\lambda)$ exists, not
depending upon either $n$, or $x^0$. Thus, 
\begin{equation}
 \partial_0 \kappa = \kappa \hat{a}_0(x^0, b) - \hat{a}_0( x^0, a) \kappa 
  = 0 .
 \end{equation}
  These  are strong assumptions. Previous experience with affine Toda
  theory \cite{BCDR} indicates they might  not lead to stronger constraints
  on the boundary condition than those which would be obtained by 
alternative means (such
  as examining directly the existence of conserved quantities). However,
the classical $r$-matrix for affine Toda field theory does not itself
depend on the Toda field. This fact represents a crucial difference.
  Substituting $\hat{a}_0$ into the above, one obtains
 \begin{equation}
  \lambda \left[\kappa(\lambda) , A_0\right]_- 
  + \left[ \kappa(\lambda) , B \right]_+ =0
 \end{equation}
  where
  $$
  B = 2\left(n\otimes {\cal F} - 
{\cal F}\otimes n\right).
  $$
  Obviously, $\kappa(\lambda) = I$ and ${\cal F} = 0$ is one of 
the  solutions;
it corresponds to the free boundary condition. To find the others for 
general $N$  is less easy and only the case $N=3$ will 
be discussed in detail here.

It is useful first to note that when $N=3$ 
$\kappa (\lambda )$ is an element of 
the group 
$SO(3)$ and must therefore have the form
\begin{equation}
\kappa_{ab}=P\delta_{ab} + Qk_ak_b +R\epsilon_{abc}k_c= 
\kappa^S_{ab}+\kappa^A_{ab}
\end{equation}
where $P,Q,R$ and the vector $k$ could depend upon $\lambda$ but not $n$ or 
$x^0$, and the $S,A$ superscripts denote the symmetric and antisymmetric 
parts, respectively. Orthogonality requires
\begin{equation}
P^2+|k|^2R^2=1\qquad |k|^2 Q^2 -R^2+2PQ=0.
\end{equation}
It is also useful to write
$$(A_0)_{ab}=\epsilon_{abc}\alpha_c \qquad B=\epsilon_{abc}\beta_c.$$
Taking the symmetric and antisymmetric parts of eq(31) yields a 
pair of equations:
\begin{equation}
\begin{array}{l}
\lambda\left(\kappa^S_{ad}\epsilon_{dbc}\alpha_c+\kappa^S_{bd}
\epsilon_{dac}\alpha_c\right) + R\left(\beta_ak_b+\beta_bk_a\right)=0 \\[2mm]
\ \ \lambda R\left(\alpha_ak_b-\alpha_bk_a\right)+\kappa^S_{ad}\epsilon_{dbc}
\beta_c -\kappa^S_{bd}\epsilon_{dac}\beta_c =0.\\
\end{array}
\end{equation}
  The vanishing trace of the first of eqs(34) requires $R\ (\beta\cdot k)=0$.
Therefore, assuming $R\ne 0$, $\beta$ must have the form
\begin{equation}
\beta =\rho \times k.
\end{equation}
The second of eqs(34) forces $\rho$ to be proportional to $\alpha$
and, without any loss of generality, $\beta$ may be set equal to
$\alpha \times k$. However, it is also necessary that
\begin{equation}
\lambda R + 2P + |k|^2 Q=0.
\end{equation}
Finally, returning to the first of eqs(34) and using (35) the other
components vanish provided
\begin{equation}
\lambda Q +R =0.
\end{equation}
Eqs(33), (36) and (37) (over-)determine $P,Q$ and $R$ to be
\begin{equation}
P=\pm\ {\lambda^2 -|k|^2\over \lambda^2 +|k|^2} \qquad Q =\pm\ {2
\over \lambda^2 +|k|^2}\qquad R=\mp\ {2\lambda\over \lambda^2 +|k|^2}.
\end{equation}
Finally, noting 
\begin{equation}
\beta =2 n\times {\cal F}\qquad \alpha =2n\times\partial_0 n,
\end{equation}
and, therefore,
\begin{equation}
n\times {\cal F}=(n\times\partial_0 n)\times k,
\end{equation}
it is straightforward to deduce
\begin{equation}
{\cal F}= -k\times \partial_0 n +(n\cdot k\times \partial_0 n) n
\end{equation}
together with the further constraint:
\begin{equation}
k\cdot\partial_0 n=0.
\end{equation}
To summarise, a consistent boundary condition requires
\begin{equation}
\partial_1n= -k\times \partial_0 n +(n\cdot k\times \partial_0 n) n
\quad \hbox{\bf and}\quad k\cdot\partial_0n=0\quad \hbox{at}\quad x=a.
\end{equation}
The vector $k$ may be chosen freely and represents extra parameters
allowed by the boundary condition. Notice, however, that with the exception
of the Neumann condition, the boundary violates the $O(3)$ symmetry 
of the model leaving a subgroup of rotations preserving $k$.

The choice $R=0$ leads  to ${\cal F}=0$ and $\kappa=I$.

 \noindent{\bf Construction of conserved quantities}
  
 The patching matrix $\kappa (\lambda )$ is an essential
ingredient in the construction of  conserved quantities for 
  the model defined on the  half-line $x\le a$ via the trace of the
generating function: 
 \begin{equation}
  U(\lambda) = T(-\infty, a; \lambda)\kappa(\lambda)T(b, \infty; \lambda).
 \end{equation}
Using the reflection properties,
 $$n(x) = n(a+b-x)\qquad \partial_1 n(x) = -\partial_1 n(a+b -x)$$ 
 $$\partial_0 n(x) = \partial_0 n(a+b-x)\qquad 
a_1(x, \lambda) = -a_1(a+b-x, -\lambda),$$  
eq(44) may be rewritten as
 \begin{equation}
  U(\lambda) = T(-\infty, a; \lambda)\kappa(\lambda) 
T^{-1}(-\infty,a; -\lambda).
  \end{equation}
  
Integrability requires the conserved quantities to be in involution
and, when the model is restricted to a half-line, this in turn implies a
compatibility relation which must be satisfied. This relation 
is expected to involve 
$\kappa (\lambda )$ and the classical $r$ matrix entering eq(14). The
 Poisson brackets: 
\begin{eqnarray}
  &&\{ T(x,y; \lambda) \ocomma T^{-1}(x,y; \mu) \} 
 = \Bigl(T(x,y; \lambda) \otimes I\Bigr) r(y,\lambda,\mu)
\left(I\otimes T^{-1}(x,y; 
 \mu)\right) \nonumber \\[2mm]
  &&\phantom{\{ T(x,y; \lambda) \ocomma T^{-1}(x,y;}
 -\left(I\otimes T^{-1}(x,y; \mu)\right)r(x,\lambda,\mu)
  \Bigl(T(x,y; \lambda) \otimes I\Bigr),
 \end{eqnarray}
 \begin{eqnarray}
&& \{T^{-1}(x,y; \lambda) \ocomma T(x,y; \mu) \} = 
  - \Bigl( T^{-1}(x,y; \lambda)\otimes I\Bigr) r(x,\lambda,\mu)\Bigl( 
  I \otimes T(x,y; \lambda)\Bigr)\nonumber\\[2mm]
  &&\phantom{ \{T^{-1}(x,y; \lambda) \ocomma T^(x,y;} + 
\Bigl(I \otimes T(x,y; \lambda)\Bigr)r(y,\lambda,\mu)\left( T^{-1}(x,y; 
 \mu)\otimes I\right), 
 \end{eqnarray}
  and
 \begin{eqnarray}
&& \{ T^{-1}(x,y; \lambda) ~\ocomma~ T^{-1}(x,y; \mu) \} =  
 \left( T^{-1}(x,y;\lambda) \otimes T^{-1}(x,y; \mu)\right)
  r(x,\lambda,\mu)\nonumber \\[2mm]
   &&\phantom{ \{ T^{-1}(x,y; \lambda) ~\ocomma~ T^{-1}(x,y;} 
-r(y,\lambda,\mu)\left(T^{-1}(x,y; \lambda) \otimes 
T^{-1}(x,y; \mu\right)) 
 \end{eqnarray}
imply the following Poisson bracket for a pair of generating functions,
as defined in eq(45) but with different parameters $\lambda$ and $\mu$,
 \begin{equation}
 \begin{array}{ll}
 \{ U(\lambda) \ocomma U(\mu) \} 
  =r(-\infty, \lambda, \mu)\Bigl(U(\lambda)\otimes U(\mu)\Bigr) 
  -\Bigl(U(\lambda)\otimes U(\mu)\Bigr)r(-\infty, \lambda, \mu) \\[2mm] 
 \  + \Bigl(U(\lambda)\otimes I\Bigr)r(-\infty,\lambda, -\mu)\Bigl(
I \otimes U(\mu)\Bigr)
  -\Bigl(I \otimes U(\mu)\Bigr)r(-\infty,\lambda, -\mu)\Bigl(U(\lambda)
\otimes I\Bigr) \\[2mm]
\   + \Bigl(T(\lambda)\otimes T(\mu)\Bigr)\Bigl[ 
  -r(a,\lambda,\mu)\, \kappa(\lambda)\otimes\kappa(\mu) + 
  \kappa(\lambda)\otimes\kappa(\mu) \, r(a,\lambda,\mu) \\[2mm]
\phantom{\Bigl(T(\lambda)\otimes} +
 \Bigl(I \otimes \kappa(\mu)\Bigr) r(a, \lambda, -\mu) 
\Bigl(\kappa(\lambda) 
 \otimes I\Bigr)
  -\Bigl(\kappa(\lambda) \otimes I\Bigr) r(a, \lambda, -\mu) 
\Bigl(I \otimes 
 \kappa(\mu)\Bigr)\Bigr]\cdot\\[2mm]
\phantom{ \{ U(\lambda) \ocomma U(\mu) \} 
  =r(-\infty, \lambda, \mu)\Bigl(U(\lambda)\otimes U(\mu)\Bigr) 
  -}
\cdot (T^{-1}(-\lambda)\otimes T^{-1}(-\mu))\\
 \end{array} 
 \end{equation}
  where $T(\lambda) = T(-\infty, a; \lambda)$.
 To establish
 \begin{equation}
 \{ trU(\lambda)~,~ trU(\mu) \} = 0
 \end{equation}
  it would be sufficient to require
 \begin{equation}
 \begin{array}{l}
   \Bigl[\kappa(\lambda)\otimes\kappa(\mu) , r(a,\lambda,\mu)\Bigr]
  +
 \Bigl(I \otimes \kappa(\mu)\Bigr) r(a, \lambda, -\mu) 
\Bigl(\kappa(\lambda) 
 \otimes I\Bigr) \\[2mm]
 \phantom{ \kappa(\lambda)\otimes\kappa(\mu) \, 
r(a,\lambda,\mu)
  -r(a,\lambda,\mu)\, }
 -\Bigl(\kappa(\lambda) \otimes I\Bigr) r(a, \lambda, -\mu) 
\Bigl(I \otimes 
 \kappa(\mu)\Bigr)=0,
 \end{array} 
 \end{equation}
 a condition reminiscent of those encountered in \cite{Skl} and \cite{BCDR}. 
  Obviously, eq(51) is satisfied 
  by $\kappa(\lambda) = I$, corresponding to the free boundary condition. 
However, it is not in general satisfied for the other possibilities,
principally because $\kappa$ is independent of $n$ while $r$ depends
explicitly on the field $n$ but not its derivatives.

Alternatively, to demonstrate eq(50) it would be sufficient to require
$$\Bigl[\kappa (\lambda ),~T^{-1}(-\lambda )T(\lambda )\Bigr]=0.$$
However, while no proof  (or disproof) of this relation is available,
it too would appear to be unlikely to be true given the 
conjectured $n$-independence of $\kappa$.

 \bigskip
 \noindent {\bf Conclusion and Discussion} 
  
  It has been argued that at the classical level the $O(N)$ nonlinear sigma 
  models remain integrable when restricted to a half-line 
  with a Neumann boundary condition. Infinitely many conserved charges
  in involution have been constructed via eq(41) and one might suppose this
result  to
be a strong indicator of integrability. However, Mourad and Sasaki \cite{MS}
have taken a different view, analysing the non-local charges for the theory
restricted to a half-line, and concluding the opposite. Their discovery of the
 absence of
non-local charges in the presence of the Neumann  condition would
appear to supply strong reasons for a lack of integrability for
even the simplest of boundary conditions. On the other hand, others
(for example, Ghoshal \cite{Gh2} and Mackay \cite{Mac}) 
have assumed the sigma models restricted to
a half-line are quantum integrable, and have conjectured quantum reflection 
matrices compatible with the reflection bootstrap. One would expect the
quantum integrability to be reflected in classical features of these
theories (although it must be borne in mind that the quantum sigma models
have a spectrum and other properties which are very far from being reflected 
in the classical degrees of freedom). 
It would seem one might be forced to conclude that the classical non-local
charges are not crucial for integrability.

It has  been shown that the $O(3)$ sigma model 
  on a half-line has a consistent Lax pair description 
not only for the Neumann condition at the boundary, but also for some 
  other suitably chosen conditions, reported in (43). 
However, it is not yet clear if these alternative boundary conditions allow
charges in involution.
  These conditions also present something of a puzzle because they do not 
appear to be derivable from a boundary potential added to the bulk Lagrangian.
The condition (41) would  be fine by itself and follow from the boundary
potential
\begin{equation}
{\cal B}={1\over 2} n\cdot k\times \partial_0 n + {\tau\over 2}(n^2-1),
\end{equation}
where $\tau$ is a Lagrange multiplier. However, the additional constraint (42)
needs to be imposed `by hand'. The significance of this is not really
clear. However, this fact and the remarks above suggest that the
Neuamnn condition is the only boundary condition compatible with
the classical Poisson structure.  
 
 \bigskip
  
 \noindent {\bf Acknowledgements}
  
  One of us (ZMS) would like to thank the Department of 
  Mathematical Sciences, University of Durham for kind hospitality, 
  the  Pao Foundation for a Fellowship, and the
  National Natural Science Foundation of China for  
  support. The other (EC) would like to thank Peter Bowcock, Patrick Dorey
and Alistair MacIntyre for comments.

 \end{document}